\documentclass[11pt]{JHEP}
\usepackage{latexsym}
\newcommand{\be}{\begin{equation}}
\newcommand{\ee}{\end{equation}}
\newcommand{\bea}{\begin{eqnarray}}
\newcommand{\eea}{\end{eqnarray}}
\newcommand{\ba}{\begin{array}}
\newcommand{\ea}{\end{array}}

\def\bbox{{\,\lower0.9pt\vbox{\hrule \hbox{\vrule height 0.2 cm
\hskip 0.2 cm \vrule height 0.2 cm}\hrule}\,}}
\newcommand{\dsl}{\pa \kern-0.5em /}

\newcommand{\nn}{\nonumber \\}
\newcommand{\tr}{{\rm tr}\,}
\newcommand{\und}{\underline}

\newcommand{\EQ}{\begin{equation}}
\newcommand{\EN}{\end{equation}}

\def\bbox{{\,\lower0.9pt\vbox{\hrule \hbox{\vrule height 0.2 cm
\hskip 0.2 cm \vrule height 0.2 cm}\hrule}\,}}

\newcommand{\pa}{\partial}

\def\tr{{\rm tr}}

\input amssym.def
\input amssym.tex
\font\mybb=msbm10 at 10pt
\def\bb#1{\hbox{\mybb#1}}
\def\bZ {\bb{Z}}

\def\bE {\bb{E}}

\def\tr{{\rm tr}}

\preprint{DAMTP-2000-132\\ hep-th/0012016}

\title{\Large\bf The Heterotic Dyonic Instanton}

\author{\bf E. ~Eyras$^{1}$, P.K. ~Townsend$^{2}$ and
M. ~Zamaklar$^{3}$\\

\vskip 24pt

{DAMTP, Centre for Mathematical Sciences\\
University of Cambridge, Wilberforce Road, \\
Cambridge CB3 0WA, U.K.\\}

\vskip 10pt

{\small 
\noindent $^1$ E-mail: e.eyras@damtp.cam.ac.uk\\
\noindent $^2$ E-mail: p.k.townsend@damtp.cam.ac.uk\\
\noindent $^3$ E-mail: m.zamaklar@damtp.cam.ac.uk\\}

}

\abstract{The static Yang-Mills-Higgs dyonic instanton is shown to have a
non-vanishing, but anti-self-dual, angular momentum 2-form with skew
eigenvalues equal to the electric charge; for large charge the angular
momentum causes the instanton to expand into a hyper-spherical shell.
A class of exact multi dyonic instantons is then found and then generalized
to a new class of 1/4 supersymmetric, non-singular, stationary, exact
solutions of the ten-dimensional supergravity/Yang-Mills theory. 
These self-gravitating dyonic instantons yield new heterotic string
solitons, to leading order in the inverse string tension.
}

\keywords{Supersymmetry, Heterotic, Dyon, Instanton, Rotation}

\begin{document}

\setcounter{equation}{0}

\section{Introduction}

The $SU(2)$ Yang-Mills (YM) instanton can be interpreted as a static
1/2-supersymmetric five-brane of the ten-dimensional (D=10) super-Yang-Mills
(SYM)
theory \cite{zizzi,pkt}. It has a generalisation to a five-brane solution of
the
effective D=10 supergravity/SYM field theory of the  heterotic string, with
an
instanton core in some $SU(2)$ subgroup of the heterotic gauge group
\cite{Strominger}. This `heterotic five-brane' can also be interpreted as
the lift
to D=10 of a self-gravitating instantonic-soliton in the N=2 truncation of
the
D=5 supergravity/SYM theory obtained by dimensional reduction \cite{GKLTT},
provided
that the $SU(2)$ subgroup of the instanton remains unbroken.

The N=2 D=5 supergravity/SYM theory can also be interpreted as the reduction
and
(consistent) truncation of a (1,0)-supersymmetric supergravity/SYM theory in
D=6
(which can itself be obtained from D=10 by reduction and truncation). The
`extra'
component of the YM field in D=6 becomes the adjoint Higgs field in D=5 and
if this
acquires an expectation value the $SU(2)$ gauge group will be spontaneously
broken
to $U(1)$. In this case, any configuration with non-zero instanton charge is
unstable against implosion to a singular field configuration. However, it
was
recently shown for the flat-space D=5 SYM-Higgs theory that an instanton
configuration can be stabilized at a non-zero radius by the inclusion of
electric
charge \cite{LT};  specifically, the energy is minimized by a {\it static}
1/4
supersymmetric configuration called a `dyonic instanton' (which has a IIA
superstring interpretation as a D0 charge distribution on two parallel
D4-branes
that is prevented from collapse by a IIA string stretched between the
D4-branes
\cite{MZ}). 

The dyonic instanton of the D=5 SYM-Higgs theory can also be
interpreted as a solution of the pure D=6 SYM theory with the charge
of the D=5 solution re-interpreted as momentum in the extra direction. As
such, it has an immediate generalisation to a 1/4 supersymmetric solution of
the
D=10 SYM theory. An obvious question is whether this D=10 SYM
configuration can serve as the core of a non-singular self-gravitating 1/4
supersymmetric `heterotic dyonic instanton' solution of the D=10
supergravity/SYM theory. In addressing this question it is crucial to 
take into account an unusual
feature of the dyonic instanton that has not hitherto been appreciated:
although it is a {\it static} solution of the D=5 SYM-Higgs theory, {\it
it has a non-zero angular momentum, proportional to the electric
charge}; inspection of the energy density shows that for
large electric charge the angular momentum causes the instanton to 
expand into a hyper-spherical shell. More precisely, if the instanton
is self-dual then the angular momentum 2-form $L$ is anti-self-dual, 
and vice versa, with skew eigenvalues equal to the electric charge. 
Although it is a surprise to find that a static
soliton may have a non-zero $L$, one might have anticipated that any
non-zero $L$
would be (anti)self-dual because this is known to be a consequence of
preservation
of supersymmetry in the context of D=5 black holes \cite{GMT}. In any case,
it
follows that a generalization of the dyonic instanton to a 1/4
supersymmetric solution of the D=10 supergravity/SYM theory should also have
non-zero angular momentum. This means that the solution we seek is not a
simple
`superposition'  of the 1/2 supersymmetric supergravity/SYM solutions having
only
instanton charge or only electric charge. However, there is another method
available, which we outline, and then exploit, below.

We shall work with the version of the supergravity/SYM theory in which the
usual 2-form potential in the supergravity multiplet is replaced by its
6-form
dual $A$ with 7-form field strength $F$ \cite{Chamseddine}. The bosonic
action in Einstein frame is
\be
{\cal S} =  \int\! d^{10}x\, \sqrt{-g}
\left(R - {1 \over 2} (\partial \phi)^2
- {1 \over 2 \cdot 7!}e^\phi F^2
- {\alpha' \over 4} e^{-\phi/2} \,  tr( G^2 ) \right)
+ {\alpha' \over 2} \int A \wedge tr( G \wedge G ) \, ,
\label{ACTION}
\ee
where $G$ is the YM 2-form field strength for the matrix-valued YM gauge
field 1-form $B$. We have chosen to write this action as it would
appear (after dualization of the 6-form gauge potential)
in an $\alpha'$ expansion of the effective action for the
heterotic string; $\alpha'$ has dimensions of length squared and
is proportional to the inverse string tension\footnote{Note that all
terms in the Lagrangian have the same dimension if all supergravity
fields are taken to be dimensionless and the YM gauge potential
is assigned the dimension of inverse length.}. Note the presence of
the $AGG$ Chern-Simons term, which plays an important role in
the self-gravitating dyonic instanton solution. In the effective
action for the heterotic string there is also a term
of the form $\alpha'\int A\wedge R\wedge R$ but this is not needed
for supersymmetry and (as in \cite{Strominger}) is actually higher-order
in $\alpha'$ for the self-gravitating dyonic instanton. Of course,
for heterotic string applications the gauge group is
$E_8\times E_8$ or $Spin(32)/\bZ_2$ but
as only an $SU(2)$ subgroup will be needed here we shall take the
gauge group to be $SU(2)$.

A dyonic instanton has a definite size, inversely proportional to the
square root of the Higgs field expectation value \cite{LT}.
As a source for the D=10 supergravity fields it may therefore be made very
diffuse
by choosing a very small expectation value for the Higgs field. For a
diffuse source
the weak field approximation is good and for this we need only the
linearized
supergravity equations. We begin by finding the solution of these linearized
equations; these linearized results serve to motivate an ansatz for the full
solution, which we then refine by demanding that it preserve 1/4
supersymmetry. It
is crucial to the success of this step that the supersymmetry
transformations of the
fermions in the supergravity multiplet be independent of the YM fields. This
is true
in D=10, but not necessarily in ${\rm D}<10$ because of the field
redefinitions
needed to put the reduced action in standard form; for this reason the D=10
starting point chosen here is the simplest.

The 1/4-supersymmetric ansatz for the supergravity fields can now be
used in the gaugino supersymmetry transformation; setting this to zero
and imposing the conditions on the supersymmetry parameter already
found from the flat space analysis yields the conditions for
preservation of 1/4 supersymmetry in the YM sector. In principle
these might depend on the supergravity fields
because the gaugino supersymmetry transformation certainly depends on
them (even in D=10). If this were the case, the dyonic instanton
equations would be modified by functions that are determined by the
supergravity field equations, but since these equations involve the YM
fields we would then be left with an intractable set of coupled
equations. Remarkably, however, the conditions for 1/4 supersymmetry
in the YM sector turn out to be {\it exactly the same as in flat-space}
(as is known to happen for 1/2 supersymmetry in the zero charge/momentum
case \cite{GKLTT}). We provide an exact multi dyonic instanton
solution of these equations based on the 't Hooft ansatz, thus
generalizing the one-instanton solution of \cite{LT}. This constitutes
a {\it known} source for the supergravity equations, which reduce to a
set Poisson equations. Remarkably, these equations can also be solved
exactly. The solution is asymptotic to a black brane solution
describing a rotating superposition of an NS-$5$-brane and
a Brinkmann wave, but with a definite value of the angular momentum that
is determined by the 5-brane charge and wave momentum.

\section{Multi Dyonic Instantons}
\label{SECTION-2}

We begin with a presentation of an explicit multi-instanton generalization
of the flat space dyonic instanton of \cite{LT}, and a determination of some
of its properties including, crucially, its angular momentum 2-form.  For
convenience we will work with the pure YM theory in the flat  D=6 spacetime
$\bE^{(1,4)}\times S^1$ with coordinates
\be
x^\mu=(x^0,x^i,x^5)\, ,\qquad (i=1,2,3,4) \, .
\ee
The YM action is 
\be
{\cal S} = -{1 \over 4} \int d^6x \, tr\left(
G_{\mu\nu}G^{\mu\nu}\right) \, ,
\ee
where $G$ is the 2-form field strength
\be
G_{\mu\nu} = \partial_\mu B_\nu - \partial_\nu B_\mu - i[B_\mu,B_\nu]
\ee
for the hermitian 1-form potential $B$. We take $B$ to  be in the
fundamental representation of $SU(2)$, so
\be
B= B^a \left({\sigma^a\over2}\right) \, .
\ee
where $\sigma^a$ $(a=1,2,3)$ are the three Pauli matrices.
The energy momentum tensor
is
\be
T_{\mu\nu} = tr\left( G_{\mu\lambda} G_{\nu}{}^\lambda \right)
- {1 \over 4}\eta_{\mu\nu}  tr\left(G_{\lambda\rho} G^{\lambda\rho}
\right)\, ,
\ee
where $\eta_{\mu\nu}$ is the Minkowski metric.
There is also an identically-conserved topological 2-form current density
\be
{\cal J}^{\mu\nu} = {1 \over 8}
\varepsilon^{\mu\nu\rho\lambda\sigma\kappa}
\, tr\left( G_{\rho\lambda} G_{\sigma\kappa} \right) \, .
\ee

Assuming that all fields are independent of $x^5$ the energy is
\be
E= {1\over2}\int d^4x\, \left[ \tr\left(G_{0i}^2\right) +
\tr\left(G_{5i}^2\right)
 + {1\over2}\tr\left(G_{ij}^2\right) + \tr G_{05}^2\right]\, .
\ee
This is subject to the Gauss law constraint
\be
{\cal D}_i G_{0i} = 0 \, ,
\label{GAUSS}
\ee
where ${\cal D}$ is the gauge-covariant derivative. Following \cite{LT}, we
rewrite
$E$ as 
\bea
E &=& {1\over2} \int d^4x\, \left[ \tr\left(G_{0i}-s'G_{5i}\right)^2 +
{1\over4}\tr\left(G_{ij}-s\tilde G_{ij}\right)^2 + \tr G_{05}^2 \right]\nn
&&\ + {s\over 4} \int d^4x\, \tr\left(G_{ij}\tilde G_{ij}\right) + s'\int
d^4x\,
\tr\left(G_{0i}G_{5i}\right)
\eea
where $\tilde G_{ij}$ is the dual field strength, and $s$ and $s'$ are
arbitrary 
signs. Since $B_i$ is independent of $x^5$ we have
\be
\tr\left(G_{0i}G_{5i}\right) = \partial_i \tr\left(B_5G_{i0}\right)\, .
\ee
It follows that, for fixed asymptotic behaviour of the fields,
\be
E \ge \big|\int \tr \left(G\wedge G\right)\big| + \big|\oint dS_i
\tr\left(B_5G_{i0}\right)\big|\, ,
\ee 
with equality when 
\be\label{inst}
G_{ij} = s{1 \over 2}\epsilon^{ijkl} G_{kl}
\ee
and 
\be\label{dyonic}
G_{5i} =  s'G_{0i}
\ee
and 
\be\label{zerofive}
G_{05}=0\, .
\ee
The energy of a solution to these equations is
\be\label{FLAT-ENERGY}
E =  4\pi^2|I| + |vq| \, .
\ee
where $I$ is the instanton number
\be
I= {1\over 32\pi^2}\int d^4x\, \varepsilon^{ijk\ell} \tr
\left(G_{ij}G_{k\ell}\right)\, ,
\ee
and $q$ is the electric charge
\be\label{electric}
q = {1\over v}\oint dS_i \tr\left(B_0 G_{i0}\right) \, .
\ee
The conditions on the YM field strength imply the following relations
between the
components of the energy-momentum tensor and the topological current
density:
\bea
T_{ij} &=& 0\nn
T_{5i} -s' T_{0i} &=& 0\, ,\nn
T_{00} - 2s' T_{05} + T_{55} &=& 0\, ,\nn
{\cal J}^{5i} +s' {\cal J}^{0i} &=& 0 \, \nn
T_{0i} + ss'{\cal J}^{0i} &=& 0\, \nn
T_{00} -T_{55} -2s{\cal J}^{05}&=& 0\, .
\eea
We also note, for future reference, that
\be
\int d^4x {\cal J}^{05} = 4\pi^2 I\, , \qquad \int d^4 x\, T^{05} =
- s' |vq|\, .
\ee
The second of these integrals is the momentum in the $5$-direction. The
other space components of the 6-momentum vanish.

As in \cite{LT}, we shall seek {\it time-independent} solutions that
minimise the
energy for fixed $I$ and $q$. The equations (\ref{dyonic}) and
(\ref{zerofive}) can then be solved by setting
\be\label{bee5}
B_5=s'B_0\, .
\ee
Because of the Gauss law constraint, $B_0$ must satisfy
\be\label{covlaplace}
{\cal D}^2 B_0 =0\, .
\ee
Given a solution of (\ref{inst}), and the boundary condition that
\be
B_0 \rightarrow v \left({\sigma^3\over2}\right)
\ee
for constant $v$, the equation (\ref{covlaplace}) has a unique non-singular
solution.
We shall consider here only those solutions of (\ref{inst}) that can be
found via the 't Hooft ansatz. This ansatz makes use of the fact that
$\bE^4$ admits two commuting sets of complex structures obeying the algebra
of the
quaternions. For one set, $\eta^a$, the corresponding K\"ahler 2-forms are
self-dual;
for the other set, $\bar\eta^a$, they are anti-self-dual. In cartesian
coordinates,
for which the metric is $\delta_{ij}$, the 2-forms $\eta^a$ obey the
following
identities
\bea\label{usefuliden}
\varepsilon^{abc} \eta^b_{ik} \eta^c_{jl} &=& -\delta_{ij}\eta^a_{kl} -
\delta_{kl}\eta^a_{ij} + \delta_{il}\eta^a_{kj} + \delta_{kj}\eta^a_{il}\,
\nn
\eta^a_{ij}\eta^a_{kl} &=& \delta_{ik}\delta_{jl} - \delta_{il}\delta_{kj} +
\varepsilon_{ijkl}\nn
\eta^a_{ik}\eta^b_{kj} &=& -\delta^{ab}\delta_{ij} + \varepsilon^{abc}
\eta^c_{ij}
\eea
The same identities are obeyed by the anti-self-dual 2-forms $\bar\eta^a$.
The 't Hooft ansatz is
\be\label{thooft}
B^a_i = \bar\eta^a_i{}^j \partial_j \log H\,
\ee
and it yields a self-dual 2-form $G$ if $H$ is harmonic on $\bE^4$. Given
that $H$ is harmonic we have
\be\label{fieldstrength}
G^a_{ij} = 2H^{-1}\bar\eta^a_{[j}{}^k \partial_{i]}\partial_k H
+4H^{-2}\left(\bar\eta^a_{[i}{}^k \partial_{j]} H -
{1\over4}\bar\eta^a_{ij}\partial_k
H\right)\partial_k H
\ee
replacing $\bar\eta^a$ by $\eta^a$ yields an anti-self-dual $G$. Note that
$G$ is self-dual for anti-self-dual K\"ahler 2-forms and {\sl vice-versa}.
If we choose $H$ to have $N$ isolated point singularies, and $H=1$ at
infinity, then we have a non-singular solution with $I=N$. The asymptotic
form of this
solution is
\be\label{asymH}
H = 1 + {\rho^2\over r^2} + {\cal O}\left( r^{-3}\right)
\ee
where $\rho^2$ is the sum of the residues of the $N$ singularities. The same
ansatz with $\eta$ in place of $\bar\eta$ yields an anti-self-dual $G$ with
$I=-N$.

Given $B_i$ of the form (\ref{thooft}) (and $H$ harmonic on $\bE^4$), the
equation (\ref{covlaplace}) has the solution
\be
B_0 = vH^{-1} \left({\sigma^3\over2}\right)\, .
\ee
When this is used in the formula (\ref{electric}) for the electric charge we
find that
\be\label{electriccharge}
q = 2\pi^2 v\rho^2\, .
\ee
For the one-instanton solution with
\be\label{onedi}
H= 1+{\rho^2/ r^2}
\ee
we recover the dyonic instanton with $I=1$ of \cite{LT}. In this case the
asymptotic form (\ref{asymH}) is exact, and $\rho$ can be interpreted as the
instanton
size. For fixed $q$ we have
\cite{LT}  
\be\label{size}
\rho = {1\over\pi} \sqrt{{q\over 2v}}\, ,
\ee
so the instanton size is not a free parameter; the collapse of the instanton
to zero size is prevented by its charge. An alternative explanation, as we
explain below, is that the collapse is prevented by angular momentum.

\section{Angular Momentum}

The angular momentum 2-form (on $\bE^4$) is
\be
L_{ij} = \int\! d^4 x\, (x^i T_{0j} -  x^j T_{0i})\, .
\ee
Noting that
\be
T_{0i} = -\partial_j \tr\left(B_0 G_{ij}\right)\, ,
\ee
and integrating by parts we deduce that
\be\label{ellij}
L_{ij} = 2\oint dS_k \left[x^j\tr\left(B_0 G_{ik}\right)  -
x^i\tr\left(B_0 G_{jk}\right)\right] -2 \int d^4x\,
\tr\left(B_0G_{ij}\right)\, .
\ee
The second term of this expression must vanish because for, say, self-dual
$G$ it
would have to be proportional to the anti-self-dual $\bar\eta^3_{ij}$, which
is
impossible unless the constant of proportionality vanishes. This can be
verified
from the expression
\be
\tr \left(B_0 G_{ij}\right) =v\left(\bar\eta^3_{[i}{}^k \delta_{j]}^\ell
-{1\over4}\bar\eta^3_{ij}\delta^{k\ell}\right)\partial_k\partial_\ell H^{-1}
\, ,
\ee
which follows from (\ref{fieldstrength}). This implies that
\be\label{GENERIC-ASYMPTOTIA}
\tr\left(B_0G_{ij}\right) \sim {2v\rho^2\over r^4} \bar\eta^3_{ij}
+ {4v\rho^2\over r^6} \left(\bar\eta^3_{\ell i} x^j - \bar\eta^3_{\ell
j}x^i\right)x^\ell + \dots
\ee
for large $r$. Using this in the first term of (\ref{ellij}) we find that
\be
L_{ij} = -2\pi^2 v\rho^2 \bar\eta^3_{ij}\, ,
\ee
which shows that $L$ is anti-self-dual. For anti-self-dual $G$ we must
replace
$\bar\eta^a$ by $\eta^a$, so that $L$ is then self-dual. Comparing with the
expression (\ref{electriccharge}) we see that
\be
L_{ij} = -  q \bar\eta^3_{ij}\, ,
\ee
which shows that $L$ is non-zero when $q$ is non-zero.

The angular momentum has a dramatic effect on the energy density ${\cal E}$.
For the
multi dyonic instanton one finds that
\be
{\cal E} \equiv T_{00} = {1\over4}\Box\left\{H^{-2}\left[\left(\partial
H\right)^2
+ v^2\right]\right\}\, ,
\ee
where $(\partial H)^2 =\sum_i(\partial_i H)(\partial_i H)$. For $H$ as in
(\ref{onedi}) this reduces to
\be
{\cal E} = {24\rho^4\over(r^2 + \rho^2)^4}\left[ 1 +
{1\over4}\left(vr\right)^2\right]\, .
\ee
Using (\ref{size}) and defining the dimensionless constant
\be
a= {vq\over 2\pi^2}\, ,
\ee
we can rewrite the energy density as
\be
{\cal E}(vr) = 12v^4 a^2 \left({2+ (vr)^2\over \left[(vr)^2 +
a\right]^4}\right)\, .
\ee
For $a<8$ this function has its maximum, and only stationary point, at
$r=0$. For
$a>8$ the stationary point at $r=0$ is a minimum and there is one other
stationary
point, a maximum, at
\be
vr= \left({a-8\over 3}\right)^{1\over2}\, .
\ee
For large $a$ the maximum occurs for
\be
vr\approx \sqrt{a\over3} \gg 1\, ,
\ee
while the ratio of ${\cal E}$ at the origin to its value at the maximum goes
to
zero as $1/a$. In other words, {\it the angular momentum has has blown up
the
spherical instanton into a hollow hyper-spherical shell}. Note that angular
momentum
can prevent the collapse of an $(n-1)$-sphere in $\bE^n$ only for {\it even}
$n$ (because rotation matrices are necessarily singular when $n$ is odd);
here we have an example of this for $n=4$.

\section{Weak field supergravity}

We now move to D=10, taking the D=10 Minkowski space coordinates to be
\be
x^M = (x^\mu,x^m)\, \qquad (m=6,7,8,9).
\ee
The bosonic equations of motion that follow from the action
(\ref{ACTION}) are, on setting $\alpha'=1$,
\bea
\partial_M \left( \sqrt{-g} \, \partial^M \phi \right)
&=& {1 \over 2 \cdot 7!} e^\phi F^{2} -  {1 \over 8} e^{-\phi/2}
\, tr( G^2 )\, ,
\nonumber \\
\partial_{M_1} \left(
 \sqrt{-g} e^\phi F^{M_1 \dots M_7} \right) &=&
-{1 \over 8} \epsilon^{M_2 \dots M_7 PQRS} \, tr(G_{PQ} G_{RS})  \, ,
\nonumber \\
{\cal D}_M \left( \sqrt{-g} e^{-\phi/2} G^{MN} \right)
&=& {1 \over 7!} \epsilon^{M_1 \dots M_7PQN} F_{M_1 \dots M_7} G_{PQ} \, ,
\\
{\cal G}_{MN} &=& 
{1 \over 2} \left( T^{(\phi)}_{MN} + T^{(F)}_{MN} + T^{(G)}_{MN}\right)\, ,
\nonumber 
\eea
where
\bea
{\cal G}_{MN} &=& R_{MN} - {1 \over 2} g_{MN} R \, ,\nonumber \\
T^{(\phi)}_{MN} &=& - {1 \over 2} g_{MN} (\partial \phi)^2
+ \partial_M \phi \partial_N \phi \, ,\nonumber \\
T^{(F)}_{MN} &=& {1 \over 6!} e^\phi
\left( F_{M P_1 \dots P_6}F_N{}^{P_1 \dots P_6} - {1 \over 14} g_{MN}
F^{2} \right)
\, ,\\
T^{(G)}_{MN} &=&  e^{-\phi/2}
\left( \, tr( G_{MP} G_N{}^P ) - {1 \over 4} g_{MN} \, tr( G^2 )\, \right)
\, .\nonumber
\eea

Given a diffuse YM source we may use a weak field approximation to
these equations. We shall begin by considering this case, for which the D=10
dilaton and 7-form field strength are small, as is the metric perturbation
$h=g-\eta$, where $\eta$ is the D=10 Minkowski metric. Under these
circumstances the linearized supergravity equations suffice, and we
can ignore any supergravity corrections to the SYM equations.
We shall further assume that the SYM fields satisfy the
dyonic instanton equations (\ref{inst}), (\ref{dyonic}),
(\ref{GAUSS}) and (\ref{zerofive}).

Given these assumptions, the linearized $A$ field equation implies the
equations
\bea
\pa_{i} F_{i056789}&= &{s\over 4} tr(G_{ij}G_{ij}) \, ,\nn
\pa_{i} F_{ij06789}&= & {ss'\over 2}
tr(G_{5i}G_{ij}) \, ,\\
\pa_{i} F_{ij56789} &= &{s\over2}
tr(G_{0i}G_{ij}) \, .\nonumber
\eea
We similarly find from the linearized dilaton equation that
\be
\Box \phi = - {1\over 8} tr(G_{ij}^2) \, ,
\ee
and from the linearized Einstein equation (in the de Donder gauge) that
\be
\ba{rclrcl}
\Box h_{00} &=& {1\over 2} tr(G_{0i}^2) + {1\over 32} tr(G_{ij}^2)  \, ,&
\Box h_{55} &=& {1\over 2} tr(G_{0i}^2) - {1\over 32} tr(G_{ij}^2)  \, ,
\\ & & \\
\Box h_{50} &=& {1\over 2}s' tr(G_{0i}^2) \, ,&
\Box h_{0i} &=&-{1\over 2} tr(G_{0j}G_{ij}) \, ,
\\ & & \\
\Box h_{5i} &=&-{1\over 2}s' tr(G_{0j}G_{ij}) \, ,&
\Box h_{ij} &=&-{3\over 32} \delta_{ij} tr(G_{kl}^2) \, ,
\\ & & \\
\Box h_{mn} &=& {1\over 32} \delta_{mn} tr(G_{ij}^2) \, .&
& &
\label{boxheqs}
\ea
\ee
\noindent For the other components of $h$ the right-hand side is
simply zero.
It is already possible to deduce that certain components of the metric
and 7-form field strength must be non-zero. We also observe that there
are some relations between these components. Specifically,
\be
F_{ij06789} = s'F_{ij56789} \, ,
\label{little-f-symmetry}
\ee
and
\bea
h_{0i} -s' h_{5i} &=& 0\, ,\nn
h_{00} - 2s' h_{50} + h_{55} &=& 0\, , \nn
h_{ij} - {3\over2} \delta_{ij} \left(h_{00}-h_{55}\right) &=& 0\, ,\nn
h_{mn} + {1\over 2}\delta_{mn}  \left(h_{00}-h_{55}\right) &=& 0\, .
\eea
In principle, these relations might only hold in the weak field
approximation but we shall assume that they hold for the full solution; this
assumption can be justified {\it a posteriori}.

These results motivate an ansatz for the full solution, which we shall
present in terms of the inverse 10-metric, written in terms of an
inverse vielbein as
\be
g^{MN} = \eta^{\und{AB}}e^M{}_{\und {A}}e^N{}_{\und{B}} \, ,
\ee
where the underlined indices are the inertial frame indices,
indicating transformation properties under the Lorentz group
$SO(1,9)$. The linearized analysis
suggests the following ansatz for the components of the
inverse vielbein $e^M{}_{\und{A}}$:
\be
e^0{}_{\underline i} = e^0{}_{\underline 5} =  e^5{}_{\underline i} =0
\ee
and
\be
\ba{rclrcl}
e^0{}_{\underline 0} &=& \sqrt{AB} \, ,&
e^5{}_{\underline 0} &=& \sqrt{A\over B} (B-1) \, ,
\\ & & \\
e^5{}_{\underline 5} &=& - \sqrt{A\over B} \, ,&
e^i{}_{\underline 0} &=& e^i{}_{\underline 5} = \, - {1 \over \sqrt{AB}}
\, E_i \, ,
\\ & & \\
e^i{}_{\underline j} &=& A^{-3/2} \, \delta^i{}_{\underline j} \, ,&
e^m{}_{\underline n} &=& \sqrt{A} \, \delta^m{}_{\underline n} \, ,
\label{vielbeins}
\ea
\ee
for some functions $A$ and $B$.
The metric then takes the form
\bea
ds^2 &=& - A^{-1}(2-B- A^2E^2)dt^2 +
2s'A^{-1}(B+A^2E^2-1) dtdx_5 \nn
&&\ +\ A^{-1}(B + A^2E^2) dx_5^2
+ 2(dt - dx_5)dx_i E_i A^2 \nn
&&\ +\ A^3 dx^i dx^j \delta_{ij}
+ A^{-1} dx^m dx^n \delta_{mn}
\label{METRIC-ANSATZ}
\eea
where $E^2=E\cdot E$. This metric is asymptotically flat if,
as $r\rightarrow \infty$,
\be
A\rightarrow 1\, ,\qquad B\rightarrow 1\, ,\qquad E_i \rightarrow 0\, .
\ee

\section{Supersymmetry}

We shall now refine our ansatz for the supergravity fields
by insisting that it preserve 1/4 supersymmetry.
To do this we must examine the supersymmetry
transformations of the fermion fields. These are the gaugino
$\lambda$, the dilatino $\chi$ and the gravitino $\psi_M$. Their
supersymmetry variations are
\bea
\delta \lambda &=& {1 \over 2\sqrt{2}} e^{-\phi/4} G_{MN}
\Gamma^{MN} \, \epsilon
\, ,\nn
\delta \chi &=&
- {1 \over 2\sqrt{2}} \Gamma^M \partial_M \phi \, \epsilon -
{1 \over 4 \sqrt{2} \, 7!} e^{\phi/2}
F_{M_1 \dots M_7} \Gamma^{M_1 \dots M_7} \, \epsilon
\, ,\\
\delta \psi_M &=& D_M \, \epsilon +{1 \over 2 \cdot 8!} e^{\phi/2}
F_{M_1 \dots M_7}
\left(3 \, \Gamma_M{}^{M_1 \dots M_7} - 7 \, \delta_M{}^{M_1}
\Gamma^{M_2 \dots M_7} \right) \, \epsilon
\, ,\nonumber 
\eea
where $\epsilon$ is a chiral spinor parameter,
\be
\Gamma^{\und {0123456789}} \epsilon = \epsilon\, ,
\ee
and $D_M\epsilon$ is its Lorentz covariant derivative,
\be
D_M \epsilon = \left( \partial_M + {1 \over 4} \omega_{M \und{AB}}
\Gamma^{\und{AB}} \right) \epsilon\, ,
\ee
where $\omega$ is the spin connection. Its non-zero components (for our
inverse
vielbein ansatz) are
\bea
\omega_{5 \underline{i5}} &=&
-{1 \over 2} B^{-1/2}A^{-3} \big[
A\partial_i B + B \partial_i A + A^2 E\cdot E \partial_i A +
-2 A^2 E_i E\cdot \partial A \nn
&&\ - {1 \over 2}A^3 \partial_i(E\cdot E)
-A^3 E\cdot \partial E_i \big]\, ,\nn
\omega_{i \underline{j5}} &=& \omega_{i \underline{j0}} =
-{1 \over 2}B^{-1/2} \big[
\delta_{ij} 3 E\cdot \partial A - E_i \partial_j A -E_j \partial_i A
+A( \partial_i E_j + \partial_j E_i) \big]\, ,\nn
\omega_{0 \underline{i5}} &=& {1 \over 2} B^{-1/2}A^{-3}
\big[ (B-1)\partial_i A - A \partial_i B - 2 A^2 E_i E\cdot \partial A
+ A^2 E\cdot E \partial_i A \nn
&&\ - {1 \over 2} A^3 \partial(E\cdot E)
-A^3 E\cdot \partial E_i \big] \, ,\nn
\omega_{0 \underline{0i}} &=& {1 \over 2} B^{-1/2}A^{-3} \big[
\partial_i A + A\partial_i B - A^2 E\cdot E \partial_i A +
2 A^2 E_i E\cdot A  \nn
&&\ + {1\over 2} A^3 \partial_i (E\cdot E) +
A^3 E\cdot \partial E_i \big] \, , \nn
\omega_{5 \underline{0i}} &=& -{1 \over 2} B^{-1/2}A^{-2} \big[
\partial_i B - A E\cdot E \partial_i A +
2 A E_i E\cdot A + {1\over 2} A^2 \partial_i (E\cdot E) \nn
&&\ +\ A^2 E\cdot \partial E_i \big]\, ,\nn
\omega_{5 \underline{ij}} &=& - \omega_{0 \underline{ij}} \ =\
A^{-2} \left( -2E_{[i} \partial_{j]} A + A\partial_{[i} E_{j]} \right)\,
,\nn
\omega_{5 \underline{05}} &=& - \omega_{0 \underline{05}} \ =\   {1\over
2}A^{-2}
E\cdot \partial A \, ,\nn
\omega_{i \underline{05}} &=& {1 \over 2}B^{-1} \partial_i B \, ,\nn
 \omega_{i \underline{jk}} &=&
{3 \over 2} A^{-1} \left( \delta_{ij} \partial_k A
- \delta_{ik} \partial_j A\right) \, ,\nn
\omega_{m \underline{in}} &=&
{1 \over 2} \delta_{mn} A^{-3} \partial_i A \, ,\nn
\omega_{m \underline{0n}} &=&\omega_{m \underline{5n}} =
-{1 \over 2}B^{-1/2}A^{-2} \delta_{mn} (E \cdot \partial A) \, .
\eea
We will also need the gamma-matrix identity
\be\label{gammas}
\Gamma^{{\und i}{\und j}} \varepsilon^{ijkl} = -2\Gamma^{\und{1234}}
\Gamma^{{\und k}{\und l}}\, .
\ee

We are now in a position to determine the conditions imposed by preservation
of 1/4
supersymmetry.  The fermion field variations must vanish when the spinor
parameter
$\epsilon$ is subject to the appropriate conditions, which are essentially
determined by the vanishing of the gaugino variation. Given our ansatz for
the
supergravity fields, it can be shown that any solution of the  {\it
flat-space}
dyonic instanton equations also solves the full supergravity-coupled YM
equations. 
Anticipating this fact, we may easily determine the conditions to be imposed
on the
$\epsilon$ such that $\delta\lambda=0$. Using the chirality of $\epsilon$
and the
identity (\ref{gammas}), one finds preservation of 1/4 supersymmetry
provided that
\be\label{cond-a}
\Gamma^{\und 0}\epsilon = - \Gamma^{\und 5} \epsilon \, ,\qquad
(1 - \Gamma^{\und{1234}})\epsilon = 0 \,.
\ee
Because of the chirality of $\epsilon$, it then follows that
\be
\Gamma^{\und{6789}} \epsilon = \epsilon \, .
\ee

We now turn to the dilatino variation. We will assume
(\ref{little-f-symmetry})
because this was suggested by the linearized analysis. Using this, and
the ansatz for the metric, in $\delta\chi=0$ we deduce that
\be
\Gamma^{\und i} \left( \partial_i \phi
+ {1 \over 2} e^{\phi/2} A^{3} F_{056789i} \Gamma^{\underline{6789} }
\right) \epsilon =0\, .
\ee
This is satisfied (without further conditions on $\epsilon$) if
\be
F_{056789i} =  2A^{-3} e^{-\phi/2} \partial_i \phi \, .
\label{cond-2}
\ee

Finally, we must consider the gravitino variation. We will
consider the time and space components separately. First, the
vanishing of the time-component of $\delta\psi_M$ yields the
equation\footnote{We use conventions such that
$2\partial_{[i}V_{j]} = \partial_i V_j - \partial_j V_i$.}
\bea
0 &=&  
\left( E_i - B^{1/2} A^{-1} \Gamma^{\und i} \right)
\left( \partial_i A - {1\over 2} A^4 e^{\phi/2} F_{056789i}
\right)\epsilon 
\nonumber \\
&& + \left(-2E_{[i} \partial_{j]}A + A\partial_{[i}E_{j]}
{\bf }- {3\over 2}e^{\phi/2} A^4 F_{056789[i}E_{j]}
+ {1 \over 2} F_{06789ij} \right)\Gamma^{\und{ij}}\, \epsilon \, .
\eea
Given (\ref{cond-2}), this condition is satisfied (without further
constraints on $\epsilon$) if
\bea\label{susycond}
A &=& e^{\phi/2} \, ,\nn
F_{056789i} &=& - \partial_i \left( A^{-4} \right)\, ,\\
F_{06789ij} &=& -2 \partial_{[i} \left(A^{-1} E_{j]} \right) \, ,\nonumber
\eea
with $F_{ij6789}$ given by (\ref{little-f-symmetry}).
Note that the restrictions on $F$ are consistent with the
Bianchi identity $dF =0$. We have still to consider the space
components of the gravitino variation. It turns out that the
$5$-component now vanishes identically while the $i$-components vanish
provided that
\be
\partial_i \epsilon + {1 \over 4} \partial_i
{\rm ln}(AB) \epsilon = 0 \, .
\ee
This is solved by setting
\be
\epsilon = (AB)^{-1/4} \epsilon_0 \, ,
\ee
for a constant chiral spinor $\epsilon_0$ satisfying the same
algebraic constraints as $\epsilon$.

To summarize, we see that preservation of 1/4 supersymmetry fixes
the dilaton and seven form field strength in terms of the
functions $A$, $C$ and $E_i$ appearing in the metric, but does not
constrain these functions. They are constrained by the equations of
motion, however, and we now turn to consider them.

\section{The multi heterotic dyonic instanton}

We now turn to the full field equations.
Consider first the YM field equation. Using  the metric ansatz
and (\ref{susycond}), one finds that the time component of this equation
reduces to the {\it flat-space} Gauss law constraint (\ref{GAUSS}).
Moreover, the space components of the YM equation are solved by any
solution of the {\it flat space} equations (\ref{inst}) and (\ref{dyonic}).
The full YM equations are
therefore solved by any solution of the {\it flat space} dyonic instanton.
Given the solution of these equations previously discussed
we have a {\it known} source for
the remaining, supergravity, equations. We now turn to these
equations, setting
\be
s=1\, ,\qquad s'=-1\, ,
\ee
for convenience. Using the metric ansatz and (\ref{susycond}) one finds that
the dilaton equation reduces to
\be
\Box \left( A^4 \right) = - {1 \over 4} \, tr( G_{ij} G_{ij} )\, .
\label{BOX-A}
\ee
The equation for the supergravity six form gauge field is now identically
satisfied except for the $06789k$-component, which yields,
\be\label{eeqn}
\pa_i \left[\pa_{[i} \left(A^3E_{j]}\right)\right]
=  - {1\over2}  \, tr( G_{0i}G_{ij} )\, ,
\ee
where we have used the self-duality of $G_{ij}$ to simplify the
right-hand side. 

Finally, we must consider the Einstein equation. As in the linearized
analysis we shall impose the de Donder gauge condition
\be
\partial_M \left(\sqrt{-\det g}g^{MN}\right) =0\, .
\ee
For our metric this is equivalent to
\be\label{gaugefix}
\partial_i (A^3 E_i)=0\, ,
\ee
so in this gauge (\ref{eeqn}) simplifies to
\be
\Box (A^3E_j) = - \tr( G_{0i}G_{ij} )\, .
\ee
Note that this equation is consistent with (\ref{gaugefix}) since
$tr(G_{0i}G_{ij})= -\partial_i tr(B_0G_{ij})$. The
Einstein equations in this gauge reduce to the single equation
\be
\Box (B + A^2E^2) = \tr (G_{0i}^2)\, .
\ee

When these equations for $A$, $B$ and $E_i$ are written in terms of the
new functions 
\be\label{effs}
f_5 = A^4\, ,\qquad f_w =B-1 + A^2E^2\, ,\qquad f_i = A^3E_i\, ,
\ee
they become the Poisson equations
\bea\label{poisson}
\Box f_5 &=& - {1 \over 4} \tr (G_{ij}^2) \, ,\nn
\Box f_w  &=& \tr (G_{0i}^2) \, ,\nn
\Box f_j &=&  - \tr( G_{0i}G_{ij}) \, .
\eea
For the multi-dyonic instanton configuration presented in section 2,
these become
\bea\label{eom-matter}
\Box f_5 &=& -{1\over4} \Box \left[H^{-2}(\partial H)^2\right]\, ,\nn
\Box f_w &=& -{v^2\over4}\Box H^{-2}\, ,\nn
\Box f_j &=& \Box\left({v\over4} H^{-2} \bar\eta^3_{jk}\partial_k
H\right)\, ,
\eea
where $(\partial H)^2 = (\partial_i H)(\partial _i H)$, which should
be solved subject to the asymptotic boundary conditions
\be
f_5 \rightarrow 1\, ,\qquad f_w \rightarrow 0\, .
\qquad f_i \rightarrow 0\, .
\ee

Given that
\be
H= 1 + \sum_\alpha {\rho_\alpha^2\over |x-x_\alpha|^2}\, ,
\ee
the equations (\ref{eom-matter}) have the following solution
\bea
f_5 &=& 1 + \alpha'\left[\sum_{i=1}^N |x-x_i|^{-2} -
{1\over4} H^{-2} (\partial H)^2\right]\, ,\nn
f_w &=& {\alpha'\over4}v^2\left(1-H^{-2}\right)\, ,\nn
f_j &=& -{\alpha'\over 4}v \bar\eta^3_{jk} \partial_k H^{-1}\, ,
\eea
where we have re-instated the dimensional parameter $\alpha'$.
The sum of poles in the expression for $f_5$ is needed to ensure
non-singularity of the metric at poles of $H$. Note also that
the gauge condition $\partial_i f_i=0$
is an identity because of the antisymmetry of the matrix $\bar\eta^3$.
For $N=1$ this solution reduces to
\bea
f_5 &=&  1 + \alpha'{(r^2 + 2 \rho^2) \over (r^2 + \rho^2)^2}\, ,\nn
f_w &=&  {\alpha'\over2} v^2\rho^2 \left( {\rho^2 + 2r^2\over (r^2 +
\rho^2)^2}\right)\, , \nn
f_j &=& -{\alpha'\over2} v\rho^2 \left(r^2 +\rho^2\right)^{-2}
\bar\eta^3_{jk} x^k\, .
\eea

The metric in terms of the functions $f_5$, $f_w$ and $f_i$ is
\bea\label{sugrasol}
ds^2 &=& f_{5}^{-{1 \over 4 }} \bigg[ -(1- f_{w})dt^2 - 2f_{w}
dtdx^{5} + (1+f_{w}) (dx^{5})^{2} \nn
&&\ +\ 2f_{i}dx^{i}(dt-dx^{5}) \bigg] +
f_{5}^{3\over4} dx^{i}dx^{j}\delta_{ij} + f_{5}^{-{1 \over 4}}
dx^{m}dx^{n}\delta_{mn} \, ,
\eea 
while the other non-zero supergravity fields are
\bea
e^{2\phi} &=& f_{5} \, ,\nn
F_{056789i} &=& - \pa_{i}(f_{5}^{-1}) \, ,\nn
F_{06789ij} &=& -2 \pa_{[i} ( f_{j]} f_5^{-{1\over 4}} ) \, .
\eea

The asymptotic form of the functions $f_5$, $f_w$ and $f_i$
determining the supergravity fields of the dyonic instanton source are
\be
f_{5} = 1+ {\alpha'|I|\over r^2} \, ,\qquad
f_{w}= {\alpha' |vq| \over 4 \pi^2 r^2} \, ,\qquad
f_{i} = \alpha' {L_{ij} x^{j} \over 4\pi^2 r^4} \, ,
\label{ASYMPTOTIC-f}
\ee
where $I$ is the instanton number, $q$ the charge of the dyonic
instanton core and $L_{ij}$ is the angular momentum.
This must yield a solution of the pure D=10 supergravity theory.
Indeed, it is essentially equivalent\footnote{After correction of a
sign error in \cite{CH}; we thank Carlos Herdeiro for discussions on
this point.} to a solution of \cite{CH} representing a rotating
superposition of an NS-5-brane with a Brinkmann wave. Note, however,
that the pure supergravity solution depends on {\it} three parameters
because the angular momentum is arbitrary. In contrast, the angular
momentum is fixed in terms of the other two charges in the
non-singular supergravity/YM solution found here.

\section{ADM energy}

One can express the energy of the solution that we have just found in
terms of an ADM-type integral over the 3-sphere at transverse spatial
infinity, where the metric takes the asymptotic form
$g_{MN}=\eta_{MN} + h_{MN}$. In {\it cartesian} coordinates, this ADM
energy is (for $\alpha'=1$)
\be
E = \oint dS_i \left[ \partial_j h_{ij} -\partial_i (h_{jj} + \hat
h)\right],
\ee
where
\be
\hat h = h_{55} + h_{mm}\, .
\ee
This differs from the standard ADM formula in that it includes the
$\hat h$ term that arises from the extra dimensions \cite{KSS}. 
Application to the metric of (\ref{sugrasol}) yields
\bea\label{asflat}
M &=& \oint dS \partial_r \left[ f_5^{3\over4} -\left(4f_5^{3\over4} +
f_5^{-{1\over4}}(1+f_w) + 4 f_d^{-{1\over4}}\right)\right]\\ \nonumber
&=& - \oint dS \left(\partial_r f_5 + \partial_r f_w\right)\\
\nonumber
&=& 4\pi^2 |I| + |vq|\, ,
\eea
exactly as in (\ref{FLAT-ENERGY}).

Note that the above result for $M$ is {\it not} equal to a multiple of
the coefficient of $h_{00}$. The reason for this is as
follows. Assuming a weak source, the ADM energy energy can be
rewritten as the bulk integral
\be
E = \int\! d^4x\, \partial_i\left[ \partial_j h_{ij} -
\partial_i (h_{jj} + \hat h)\right]\, .
\ee
Use of the linearized version of the Einstein equation $G_{MN}=
{1\over2}T_{MN}$ allows us to rewrite this as
\be\label{altM}
E= -{1\over7}\int\! d^4x \left[ 8\Box h_{00} + T_{ii} + \hat T\right]\, ,
\ee
where 
\be
\hat T = T_{55} + T_{mm}\, .
\ee
{}From the observation that
\be
\int\! d^4x\, T_{ii} = \int\! d^4x\, \partial_j \left(x^i T_{ij}\right)
= \oint dS_j \left(x^i T_{ij}\right)\, ,
\ee
we deduce that the integral of $T_{ii}$ vanishes provided that it
falls off sufficiently fast near transverse spatial infinity. If it
were not for the $\hat T$ term we could then recast the integral on
the the right hand side of (\ref{altM}) as a surface integral
involving only $h_{00}$; this is why the mass of a particle-like
object can always be read off from the $g_{00}$ component of the metric.
This is not true of $p$-brane solutions with
$p>0$, as pointed out in \cite{Lu} for a class of non-extremal brane
solutions preserving 1/2 supersymmetry. However, it {\it is} possible to
express the $\hat T$ term as a surface integral, with the result that
\be
E= -{3\over2} \oint dS_i\, \partial_i (h_{00} -{1\over3}\hat h)\, .
\ee
Application of this formula to the solution (\ref{sugrasol}) again
yields the result (\ref{asflat}).


\section{Conclusions}

We have found a new exact 1/4 supersymmetric stationary solution of the D=10
supergravity/SYM theory that is asymptotic to a rotating
superposition of a black five-brane and a Brinkmann wave of the pure
D=10 supergravity theory. However, the interior is `filled in' with a
YM multi dyonic instanton core in such a way that the spacetime topology is
the
same as that of D=10 Minkowski space (or a toroidal compactification
of it), without singularities.  We have called this new solution the
`heterotic dyonic instanton' because it generalizes the 1/2 supersymmetric
heterotic five-brane solution of \cite{Strominger}, which was proposed
as a solution of the heterotic string to leading order in an $\alpha'$
expansion. The generalization found here
is two-fold: firstly, we have found an exact self-gravitating 5-brane
solution with a {\it multi-instanton} core and, secondly, we have
generalized this to the multi dyonic instanton. Of course, the D=10
supergravity/SYM theory serves equally as an effective field theory for
the Type I superstring, and our solution can therefore be viewed
as an approximate solution to either the heterotic or the Type I
superstring. 

We have also uncovered some unusual features of the flat-space dyonic
instanton, notably that it carries angular momentum despite the fact
that all fields are time-independent, and that the instanton expands
to a hyper-spherical shell for large angular momentum (corresponding
to large electric charge). Because of this angular momentum, the 
self-gravitating dyonic instanton is not a static solution of 
the supergravity/SYM
equations but rather a {\it stationary} one. For the `intersecting
black brane' solution of pure supergravity to which this
solution is asymptotic the angular momentum is a free
parameter, and may be set to zero. In contrast, the angular momentum
of the self-gravitating dyonic instanton is not a free parameter, and
it is necessarily non-zero. Thus, {\it the requirement that the
core of the supergravity solution be `filled in' by a non-singular
YM configuration fixes an otherwise free parameter}.

One surprise, although not one without precedent for simpler
1/2 supersymmetric supergravity/SYM solutions, is that the
{\it flat space} dyonic instanton solution continues to solve the SYM
equations even after coupling to supergravity. This feature means that
the Laplace equations that one must solve to find the intersecting
black brane solution with the same asymptotic behaviour are replaced
by Poisson equations with solutions that exist and are uniquely
determined (given the requirement of non-singularity) by this
asymptotic behaviour. The coupling to gravity thereby preserves
both the solution and its BPS nature in the simplest way
imaginable. This may have implications for brane world scenarios
because it provides strong evidence that BPS configurations that
are initially defined on a brane, without gravity, will survive the
coupling to gravity. It would be interesting to see whether a general
statement along these lines could be proved.


\section*{Acknowledgments}

We would like to thank C.~Herdeiro, D.~Mateos, G.~Papadopoulos and
K.~Peeters for
useful discussions. The work of E.E. is supported by the European Community
program {\it Human Potential} under the contract HPMF-CT-1999-00018.
M.Z. is supported by the St. John's college
Benefactor's scholarship.
This work is also partially supported by the PPARC grant
PPA/G/S/1998/00613.


\end{document}